\documentclass{icrc}

\usepackage{times}
\usepackage{graphicx}

\def\gray{$\gamma$-ray\ }
\def\grays{$\gamma$-rays\ }

\begin{document}

\title{A 3D time-dependent model for Galactic cosmic rays and \grays }
\author[1]{A. W. Strong}
\affil[1]{Max-Planck-Institut f\"ur extraterrestrische Physik, 
Postfach 1312, 85741 Garching, Germany}
\author[2,3]{I. V. Moskalenko}
\affil[2]{NRC-NASA/Goddard Space Flight Center, Code 660, Greenbelt, MD 20771, U.S.A.}
\affil[3]{Institute of Nuclear
   Physics, M.\ V.\ Lomonosov Moscow State University, 119 899 Moscow, Russia}

\correspondence{aws@mpe.mpg.de}

\firstpage{1}
\pubyear{2001}


\maketitle

\begin{abstract}

In studies of cosmic-ray (CR) propagation and diffuse continuum \gray\
emission from the Galaxy it has usually been assumed that the source
function can be taken as smooth and time-independent. However,
especially  for electrons at high energies where energy losses are
rapid, the effect of the stochastic nature of the sources becomes
apparent and indeed has been invoked to explain the GeV excess in the
diffuse emission observed by EGRET.  In order to address this problem
in detail a model with explicit time-dependence and a stochastic SNR
population has been developed, which follows the propagation in three
dimensions.  The results indicate that although the inhomogeneities
are large they are insufficient to easily explain the GeV
excess. However the fluctuations should show up in the gamma-ray
distribution at high energies and this should be observable with
GLAST. Estimates of the TeV continuum emission from the plane are
consistent with the Whipple upper limit.
\end{abstract}

\section{Introduction}
 
The diffuse continuum \gray emission from the Galaxy measured by EGRET
and COMPTEL has been the subject of many studies relating to
CR origin and propagation. Usually it has been assumed that
the source function can be taken as smooth and time-independent, an
approximation justified by the long residence time of CR in
the Galaxy, but which ignores the stochastic nature of the sources.
 
One motivation for studying the high-energy electrons is the
observation of the $>1$ GeV excess in the EGRET spectrum of the
Galactic emission, which has been proposed to originate in
inverse-Compton emission from a hard electron spectrum; this
hypothesis can only be reconciled with the local directly-observed
steep electron spectrum if there are large spatial variations which
make the spectrum in our local region unrepresentative of the
large-scale average.  \citet{SMR00} presented a study of
diffuse \grays\ based on a smooth time-independent 2D model. The new
3D model and its application to nuclei is presented in Strong \&
Moskalenko ``New developments in the GALPROP CR propagation \linebreak[4] model'',
these proceedings (hereineafter Paper I). Here we concentrate on
electrons and diffuse Galactic \grays.

\section{Effect of SNR on the electron spectrum} 

The current version of the model is described in Paper I.  The main
enhancement is the inclusion of stochastic SNR events as sources of
cosmic rays. The SNR are characterized by the mean time $t_{SNR}$
between events in a 1 kpc$^3$  unit volume, and the  time $t_{CR}$
during which an SNR actively produces CR.  For high-energy electrons
(TeV) which lose energy on timescales of $10^5$ years the effect is a
very inhomogeneous distribution. The amplitude of the fluctuations
depends  on these two parameters, both of  which are  poorly known.
$t_{SNR}$ is adjusted to be consistent with  estimates of the SNR
rate (e.g., Dragicevich et al. 1999).

We consider a model with a hard electron injection spectrum (index
1.8) as used by \citet{SMR00} following the suggestion of \citet{pohl}
that fluctuations could allow such a spectrum to be
consistent both with the GeV \gray excess and the locally observed
electron spectrum.  

\begin{figure*}[!t]
\hskip -7mm
\includegraphics[height=.3\textheight]{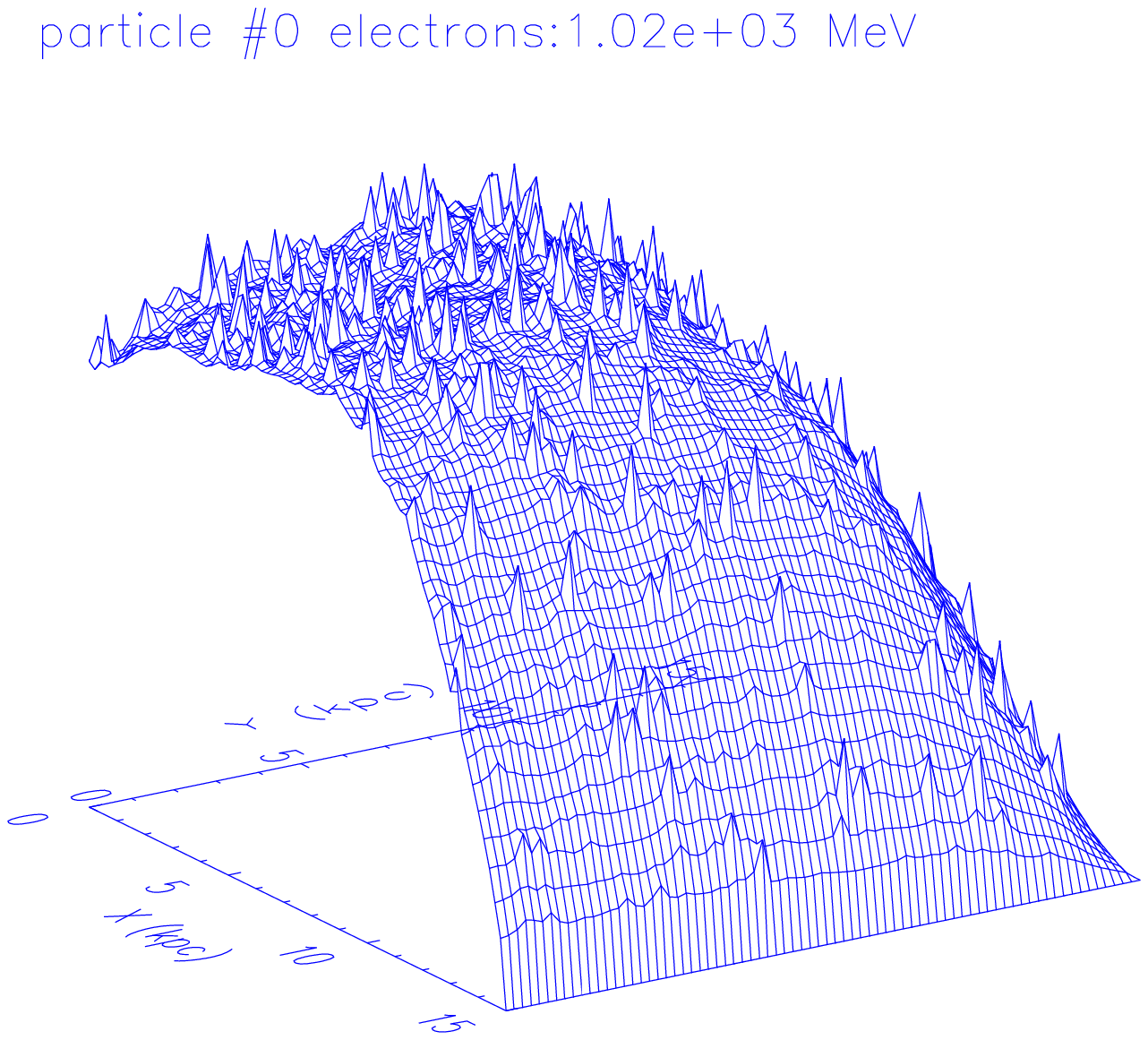}
\hskip -7mm
\includegraphics[height=.3\textheight]{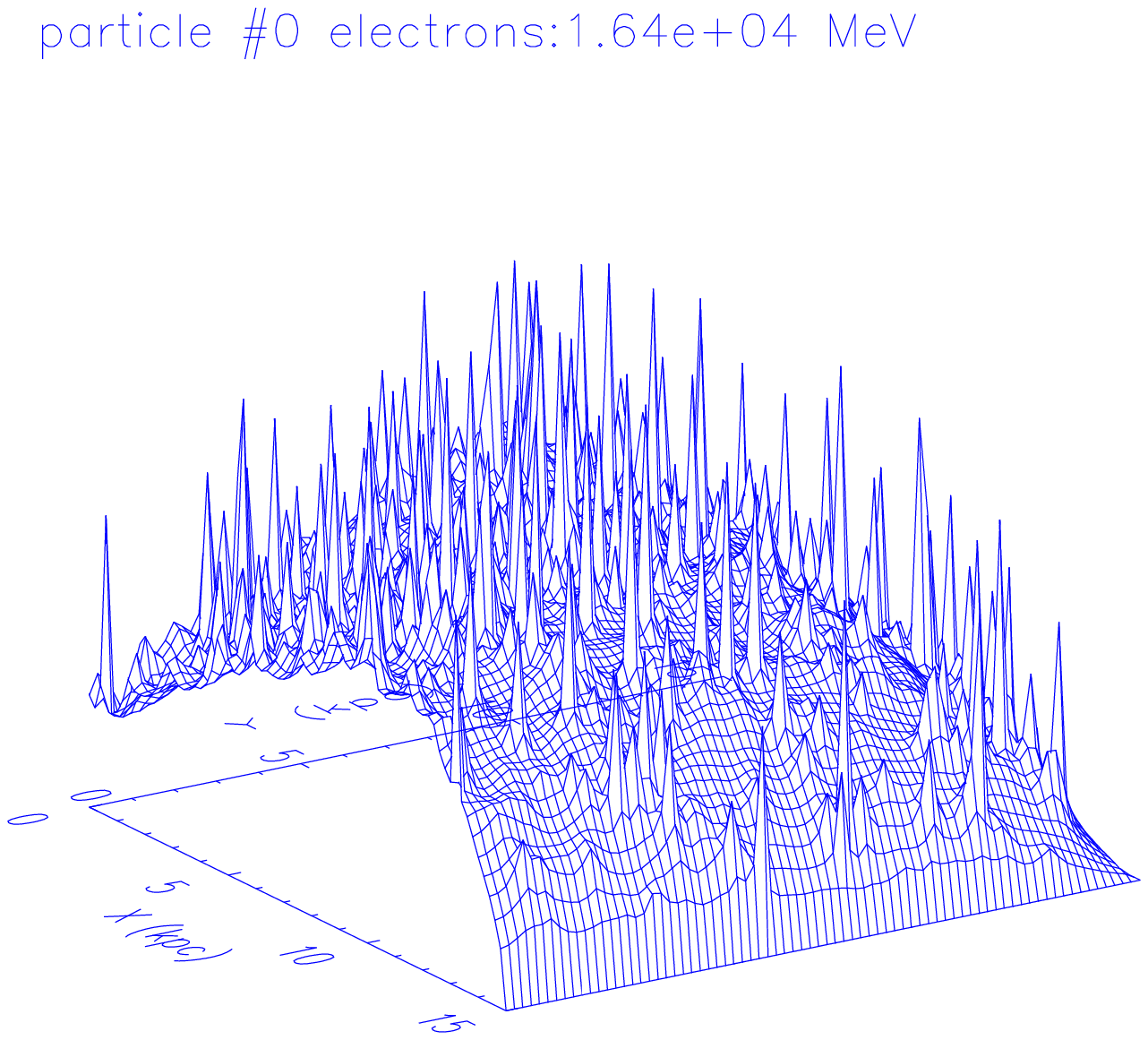}
\centerline{\includegraphics[height=.3\textheight]{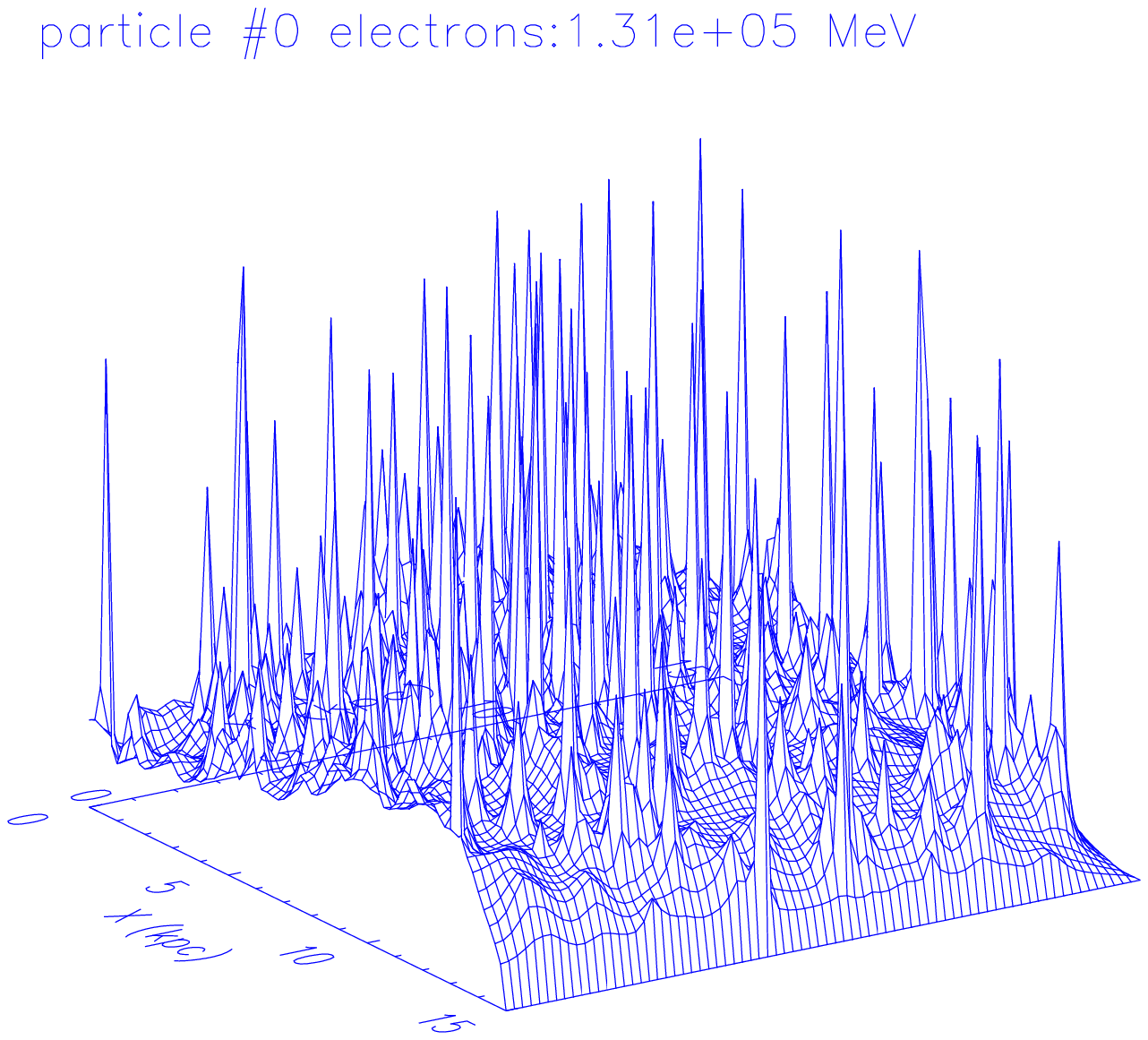}}
\caption{Distribution of 1, 16, and 130 GeV electrons at $z=0$.}
\end{figure*}
 
Figs.\ 1 and 2 show the distribution of electrons
with 1, 16, 130 and 1000 GeV, for $t_{SNR} = 10^4$  years
(corresponding to a ``standard'' Galactic SN rate 3/century). At GeV
energies the distribution shows only small fluctuations, the particle
density being dominated by the long storage times. At higher energies
the losses increase and the fluctuations become significant as the
individual SNR events leave their imprint on the distribution. The TeV
electron distribution is quite inhomogeneous,  but still none of the
spectra around $R = R_\odot$ resemble even remotely  that observed
locally (Fig.\ 2).  For  $t_{SNR} =  10^5$ years (Galactic SN rate
0.3/century)  the distribution  above 100 GeV is even more
inhomogeneous and the spectrum fluctuates even more (Fig.\ 3).  Some of
the spectra resemble that observed locally within a factor of a few,
although still none is fully compatible with the local spectrum.

We conclude that  the ``hard electron spectrum''  hypothesis for the
EGRET \gray\ excess would require an SN rate much lower than standard,
with correspondingly large power requirements for acceleration of
electrons per SNR.   It is possible that the rate of  {\it
CR-producing} SNR could  be lower than that of all SNR, so that a
sufficiently low rate could be possible, but this is unlikely in view
of the power requirements. 
The CR luminosity (electrons + nuclei) in this model is $2\times10^{41}$
erg s$^{-1}$; for 3 SNR/century we require $2\times10^{50}$ erg/SNR in CR
which is plausible (for $10^{51}$ erg and 20\% acceleration
efficiency: see Fields et al. 2001 for discussion) but for 0.3
SNR/century we need $2\times10^{51}$ erg/SNR in CR which would require
quite another class of objects.\footnote{ We note that {\it hypernovae}
may have energies of $>10^{52}$ erg \citep{nomoto}.}
It would also result in large fluctuations
in the \gray\ emission which are not evident in the EGRET data
(although this aspect should be investigated in detail).

Our conclusion differs from that of Pohl and Esposito \linebreak[4] (1998), who
found that a hard electron spectrum model is consistent with
observations considering the fluctuations, but they included however a
dispersion in the electron injection spectral index  which increases
the variations further; however our tests with such a dispersion do
not suggest that it eases the problem much since a large number of SNR
still contribute to the local electron spectrum at 1 TeV.

\begin{figure}[!t]
\hskip -7mm
\includegraphics[height=.3\textheight]{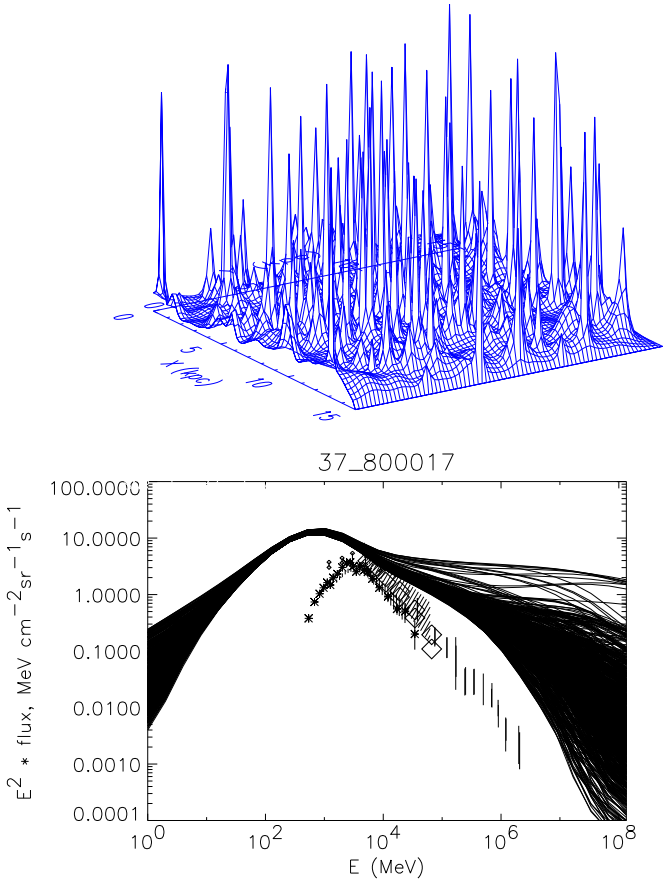}
\includegraphics[height=.25\textheight]{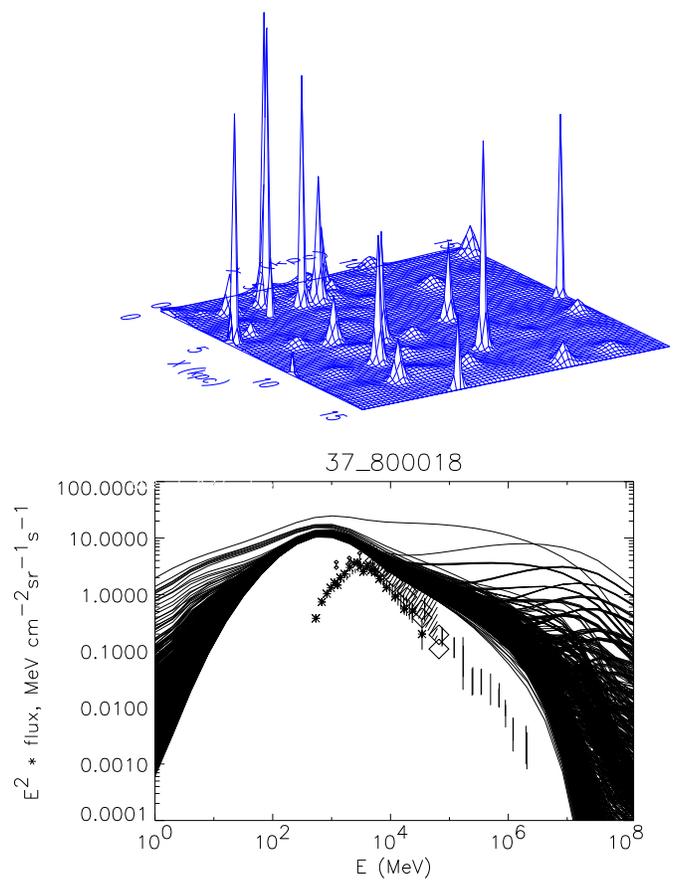}
\caption{Distribution of 1 TeV electrons at $z=0$ (top) and spectral variations 
in $4< R < 10$ kpc (bottom) for $t_{SNR} = 10^4$ yr. 
Data points: locally measured electron spectra; 
for references see \citet{SMR00}, with additional data from \citet{kobayashi}.}
\end{figure}
 
\begin{figure}[!t]
\hskip -7mm
\includegraphics[height=.3\textheight]{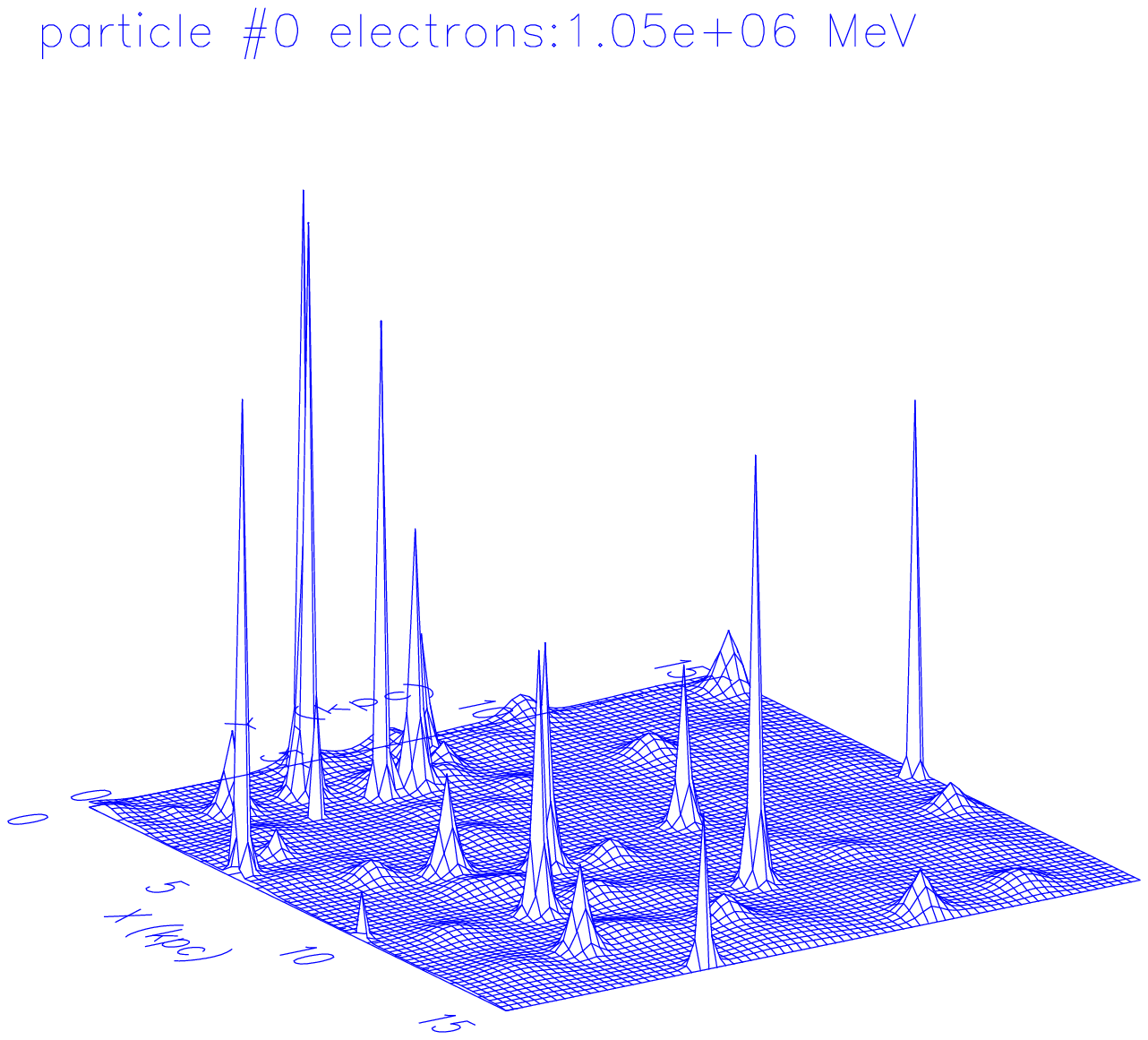}
\includegraphics[height=.25\textheight]{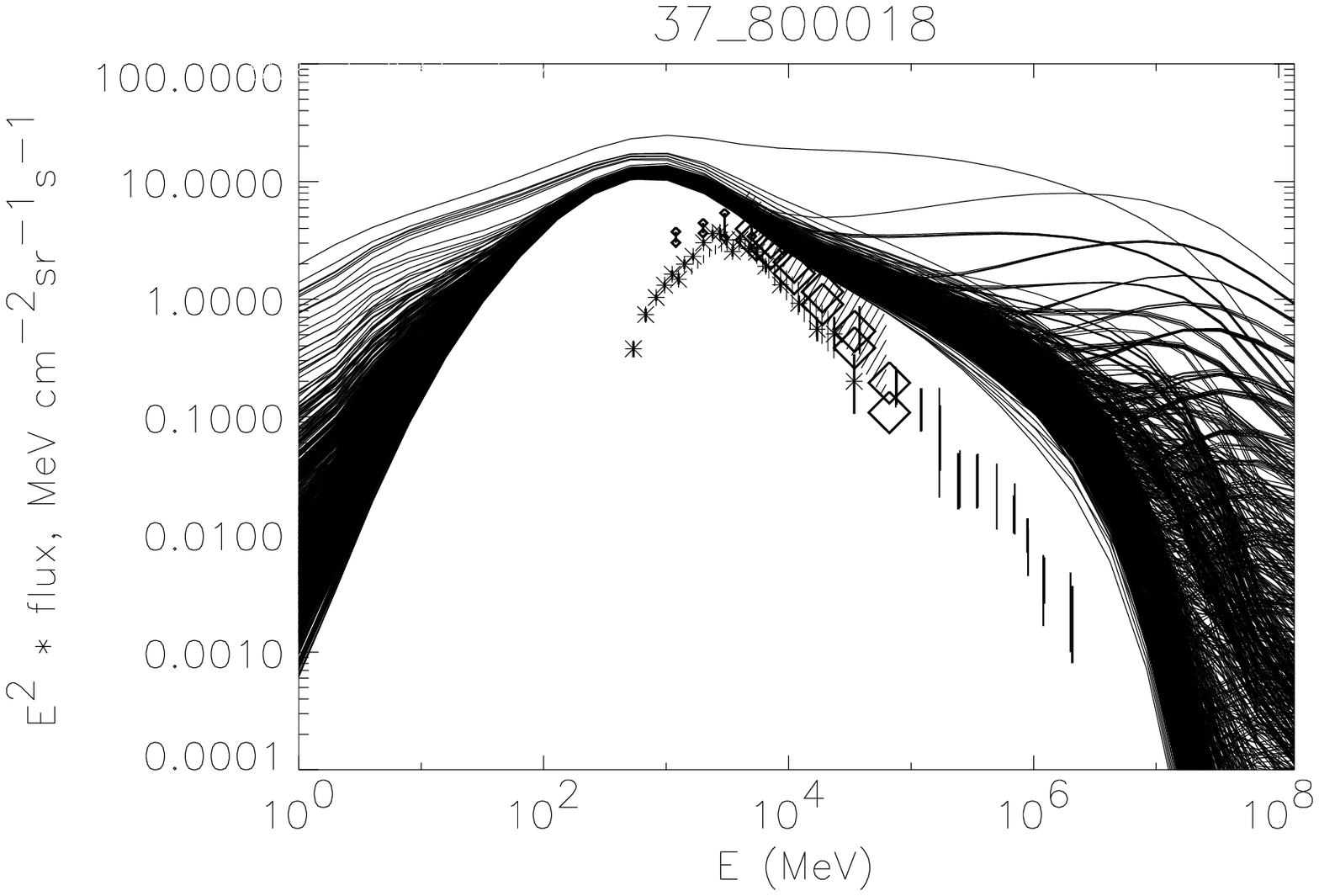}
\caption{ Distribution of 1 TeV electrons at $z=0$ (top) and spectral 
variations in $4 < R <10$ kpc (bottom) for  $t_{SNR} = 10^5$ yr. Data as Fig.\ 2.}
\end{figure}

\subsection{Effect on the \gray\ sky}

The inverse-Compton emission becomes increasingly clumpy at high
energies due to the effect of SNR, even for the standard SNR rate, as
shown in longitude distributions obtained from our model (Fig.\ 4). The
effect is already visible at  1~GeV and will be an important signature
for the  GLAST \gray\ observatory, which will measure up to 100 GeV.

\section{ Galactic diffuse TeV \grays}
Recently observations of the Galactic plane (around $l = 40^\circ$) have
been reported by the Whipple Observatory \citep{lebohec}, which
place limits on the $>500$ GeV intensity. We have extended our
predicted spectrum for a hard electron injection spectrum  to the TeV
range (Fig.\ 5). Since the maximum energy of accelerated electrons is
unknown we consider the extreme case of 100 TeV.   

Even for this case
the predicted spectrum from inverse-Compton emission is compatible
with the Whipple upper limit, and a lower  cutoff energy will be also
consistent with Whipple; it is clear that an improved limit would
quickly provide a critical test for  models with hard electron
injection spectra.  The SNR shock-acceleration models of \citet{baring}
suggest a cutoff around 1 TeV which imply a cutoff in the
\grays\ around 10 GeV, in which case the predicted intensities are well
below the Whipple limit and detection of TeV diffuse emission will be
difficult.

\begin{figure*}[!t]
\hskip -2mm
\includegraphics[height=.37\textheight]{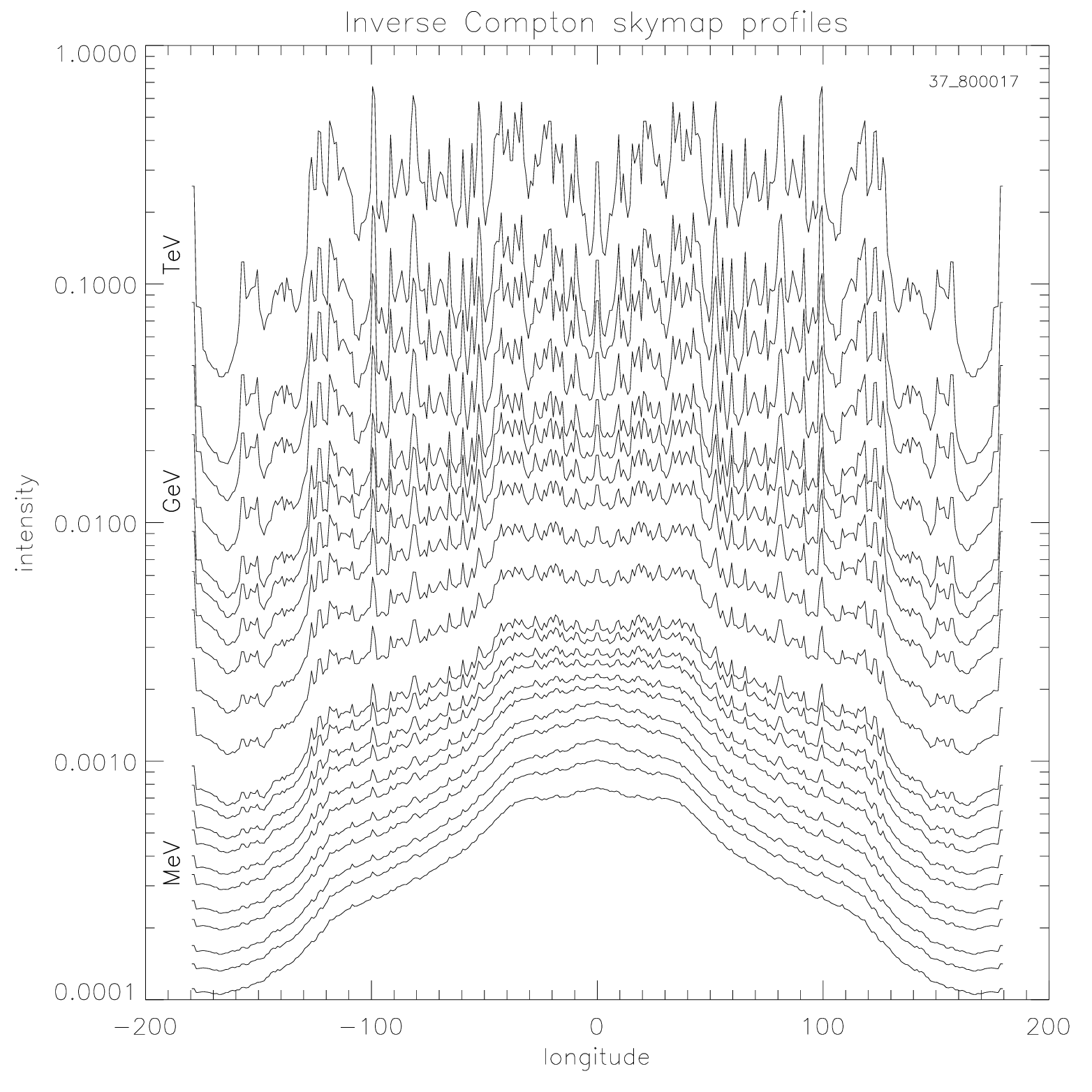}
\includegraphics[height=.4\textheight]{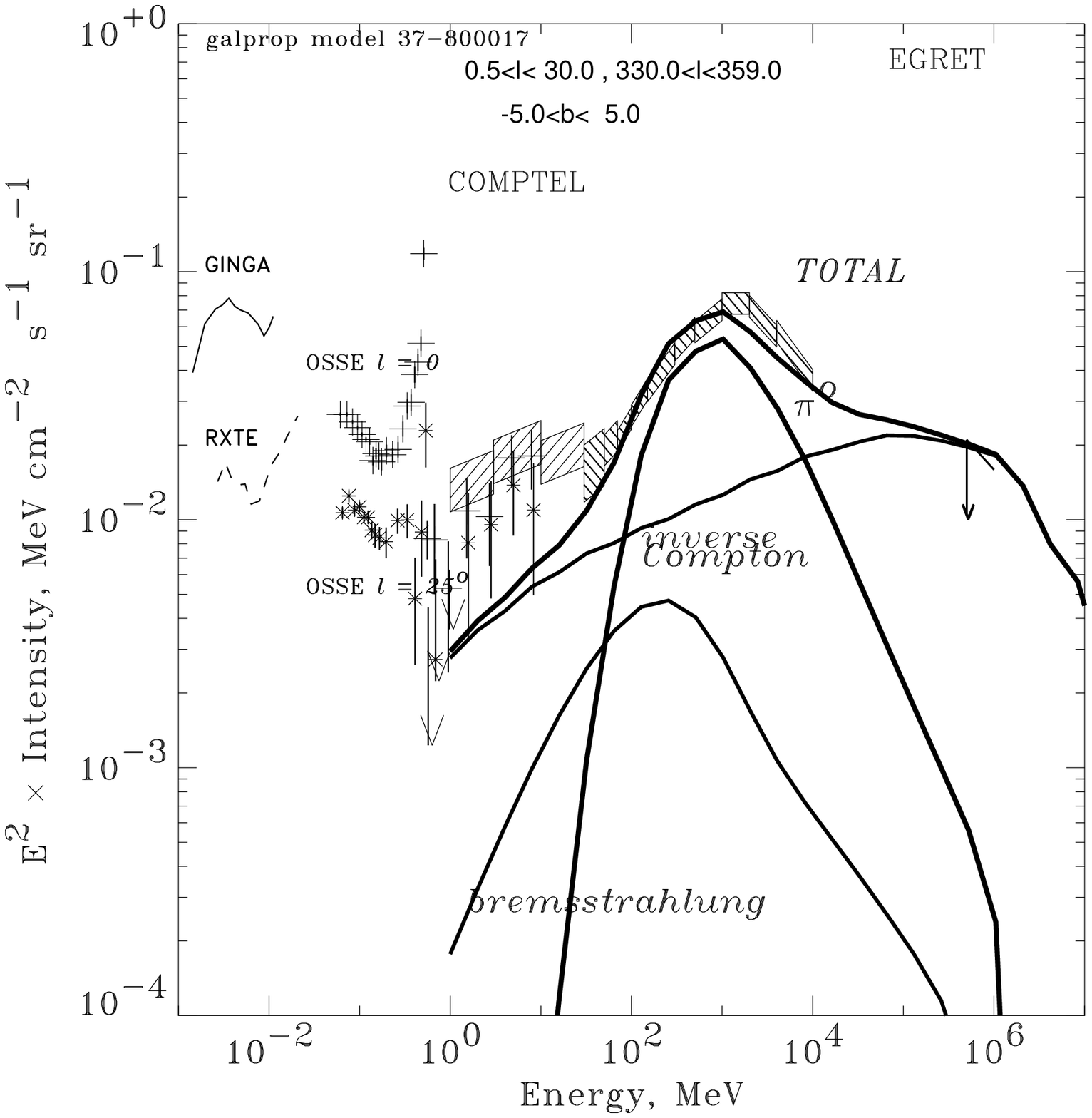}
\begin{minipage}[t]{85mm}
\caption{Inverse Compton \gray longitude distributions for 
a model with $t_{SNR} = 10^4$ yr and hard electron injection spectrum 
(index 1.8) at $b=0$ for \gray\ 
energies from 1 MeV (bottom) to 1 TeV (top). }
\end{minipage}
\hfill
\begin{minipage}[t]{85mm}
\caption{ Spectrum of the inner Galaxy ($330^\circ<l<30^\circ$, $|b|<5^\circ$) for 
a model with $t_{SNR} = 10^4$ yr and hard electron injection spectrum 
(index 1.8) and cutoff at 100 TeV. Data: EGRET \citep{SM96}, 
COMPTEL \citep{strong}, OSSE \citep{kinzer}, Whipple 
\citep{lebohec}. Note that the Whipple upper limit 
(plotted at $>500$ GeV) is for $l=40^\circ$.}
\end{minipage}
\end{figure*}

\begin{acknowledgements}
IVM acknowledges support from the NRC/NAS Research Associateship Program.
\end{acknowledgements}

\end{document}